\documentclass[aps,pra,twocolumn,showpacs]{revtex4}
\usepackage{amsmath,amsthm,amssymb}
\usepackage{graphics}

\newcommand{\ket}[1]{{|#1\rangle}}
\newcommand{\bracket}[2]{{\langle{#1}|{#2}\rangle}}
\newcommand{\ketbra}[2]{{|{#1}\rangle\!\langle{#2}|}}
\newcommand{\proj}[1]{{\ketbra{#1}{#1}}}

\newcommand{\traux}{{\tr_\mathrm{aux}}}
\renewcommand{\H}{{\mathcal H}}
\newcommand{\Haux}{{\H_\mathrm{aux}}}
\newcommand{\M}{{\mathcal M}}
\newcommand{\trnorm}[1]{{\|{#1}\|}}
\newcommand{\wcd}[2]{{\mathcal D({#1},{#2})}}

\DeclareMathOperator{\tr}{tr}
\newtheorem{theorem}{Theorem}

\begin{document}
\title{Physical Purification of Quantum States}
\author{M. Kleinmann}\email{kleinmann@thphy.uni-duesseldorf.de}
\author{H. Kampermann}
\author{T. Meyer}
\author{D. Bru\ss}
\affiliation{Institut f\"ur Theoretische Physik,
             Heinrich-Heine-Universit\"at D\"usseldorf,
             D-40225 D\"usseldorf,
             Germany
            }
\date{\today}
\begin{abstract}
We introduce the concept of a physical process that purifies a mixed quantum 
 state, taken from a set of states, and investigate the conditions under which 
 such a purification map exists.
Here, a purification of a mixed quantum state is a pure state in a 
 higher-dimensional Hilbert space, the reduced density matrix of which is 
 identical to the original state.
We characterize all sets of mixed quantum states, for which perfect 
 purification is possible.
Surprisingly, some sets of two non-commuting states are among them.
Furthermore, we investigate the possibility of performing an imperfect 
 purification.
\end{abstract}
\maketitle

\section{Introduction}
A fundamental entity in quantum mechanics and quantum information is a mixed 
 quantum state.
A mixed quantum state can be either understood as a statistical mixture of pure 
 quantum states, or as being part of a higher-dimensional, pure state -- a {\em 
 purification} of the mixed state.
Formally, given the decomposition $\rho=\sum_ip_i\proj{\chi_i}$, where 
 $p_i\geq 0$ and $\sum_ip_i=1$, an example for a purification of $\rho$ is 
 given by $\ket{\psi} = \sum_i\sqrt{p_i}\ket{\chi_i}\ket{a_i}$, with the 
 auxiliary states $\ket{a_i}$ being mutually orthogonal.
This abstract point of view was, so far, the main impetus for discussing 
 purifications of a single, {\em known} quantum state \cite{Hughston:1993PLA, 
 Bassi:2003PLA}.

In this paper, we consider the purification of an {\em unknown} quantum state.
More precisely, we introduce the fundamental question whether there exists a 
 {\em physical process} (i.e. a completely positive map) that takes any state 
 of a given set to one of its purifications.
(We remind the reader for clarity that there exists a different notion of 
 ``purification'' in the literature, referring to the process of performing 
 operations on several identical copies of a given state, such that the purity 
 of some of them is increased; a typical application is entanglement 
 distillation.)
Our aim is to characterize all sets of states for which a purifying map exists.
The existence of such a process implies a non-trivial physical equivalence 
 between certain sets of mixed and pure quantum states.

Let us introduce our concepts and outline the structure of this paper.
As already pointed out above, a purification of a mixed state has to satisfy 
 two characteristic properties: first, it has to be pure, and second, tracing 
 out the auxiliary system has to yield back the original state.
We call the second property {\em faithfulness} and name a process a {\em 
 perfect purifier} for a mixed state, when the output achieves both properties.
It is straightforward to prove that the linearity of quantum mechanics does not 
 allow the existence of a perfect purifier for a completely unknown quantum 
 state, i.e. a state taken from the set of {\em all} states.
However, will dropping the condition of faithfulness {\em or} the one of purity
  allow non-trivial purification processes for an unknown quantum state?
It will be shown in Theorem~\ref{t9167} that this is not the case.
Consequently, in Section~\ref{s4641} we will restrict the set of possible input 
 states, and investigate the properties of purifying maps acting on the most 
 simple non-trivial set, namely a set of only two mixed states.
While keeping the condition of purity, we will find that the deviation from 
 perfect faithfulness depends on a purely geometric quantity of the two inputs.
This result will allow us to derive lower and upper bounds on the achievable 
 faithfulness.
Since these bounds do not exclude perfect faithfulness for certain pairs of 
 states, we then in Section~\ref{s24871} proceed to investigate the existence 
 of a perfect purifier in general.
Theorem~\ref{t21} completely characterizes all sets of states that can be 
 purified perfectly.
Finally, we will provide an operational test for a given pair of states that 
 allows to check whether a physical purification is possible.

\section{The general purification task}
In the following we will denote by $\M$ a given set of mixed states, 
 represented by density operators that act on a finite-dimensional Hilbert 
 space $\H$.
The elements $\rho_i\in \M$ are allowed to have unbalanced {\em a priori} 
 probabilities $\eta_i> 0$, satisfying $\sum_i \eta_i= 1$.
We consider deterministic physical processes represented by completely positive 
 and trace preserving \footnote{
  The output of a probabilistic process $\Lambda_\mathrm{p}$ with success rate 
   $\tr \Lambda_\mathrm{p}[\rho]$ is (for our purposes) physically equivalent 
   to a deterministic process $\Lambda_\mathrm{d}\colon \rho\mapsto 
   \Lambda_\mathrm{p}[\rho]+ (1- \tr \Lambda_\mathrm{p}[\rho]) \proj\varphi$, 
   with $\proj\varphi$ being orthogonal to all $\Lambda_\mathrm{p}[\rho]$.
 }
 linear maps $\Lambda$ that take any density operator acting on $\H$ to a 
 density operator acting on $\H\otimes \Haux$, where $\Haux$ denotes an 
 auxiliary space of unspecified dimension.
We refer to such a physical process as a {\em perfect purifier} if for each 
 $\rho_i\in \M$, the output $\Lambda[\rho_i]$ is pure as well as faithful, i.e. 
 $\traux\Lambda[\rho_i]= \rho_i$.
If these conditions are not met, we will measure the average output purity by 
 $\mathsf p= \sum_i \eta_i \tr \Lambda[\rho_i]^2$ and the average faithfulness 
 by $\mathsf f= 1- \sum_i \eta_i \trnorm{\rho_i- \traux\Lambda[\rho_i]}$.
Here, $\trnorm{\rho-\sigma}= \frac 1 2 \tr |\rho-\sigma|$ denotes the trace 
 distance, where $|A|= \sqrt{A^\dag A}$.
The trace distance is a good measure for the distinguishability of two states 
 as it vanishes for identical states and is equal to one for orthogonal states.
In particular the success probability for the minimum error discrimination 
 procedure \cite{Helstrom:1976, Herzog:2004PRA} of two states having equal {\em 
 a priori} probability depends linearly on the trace distance of the states.
--
We call any deterministic process a {\em purifier} of $\M$, if it does not 
 decrease the average purity of $\M$.

For the universal case where the set $\M$ contains all possible density 
 operators acting on a given Hilbert space, neither relaxing the condition of 
 purity nor relaxing the condition of faithfulness allows non-trivial 
 purifiers:
\begin{theorem}\label{t9167}
 (i)~Any universal purifier with perfect output purity is a constant map.
 (ii)~A universal purifier with perfect faithfulness does not increase the 
  purity of any state.
\end{theorem}
\begin{proof}
 We prove \textit{(i)} by contradiction.
 Suppose there exists a purifier $\Lambda$ such that $\Lambda[\rho]$ is pure 
  for any state $\rho$, and with the property that at least for two states 
  $\rho_1$ and $\rho_2$, $\Lambda[\rho_1]\ne \Lambda[\rho_2]$ holds.
 But for the state $\rho_3= (\rho_1+ \rho_2)/2$, the purity of 
  $\Lambda[\rho_3]= (\Lambda[\rho_1]+ \Lambda[\rho_2])/2$ requires 
  $\Lambda[\rho_1]= \Lambda[\rho_2]$.

 Proof of statement~\textit{(ii)}: perfect faithfulness of a universal purifier 
  requires that {\em any} pure state $\proj\phi$ is mapped onto the state 
  $\proj\phi\otimes \sigma_\phi$ for some state $\sigma_\phi$ acting on 
  $\Haux$.
 For any state $\rho$ we find with the spectral decomposition $\rho= \sum 
  p_i\proj{\lambda_i}$ that due to linearity $\tr \Lambda[\rho]^2= \tr(\sum_i 
  p_i \proj{\lambda_i}\otimes \sigma_{\lambda_i})^2= \sum_i p_i^2 \tr 
  \sigma_{\lambda_i}^2\le \sum_i p_i^2= \tr \rho^2$, i.e. no state can become 
  purer by the action of $\Lambda$.
\end{proof}

Let us mention that there is some similarity of the arguments given in the 
 proof above with the no-cloning theorem \cite{Wootters:1982Nat, Dieks:1982PLA, 
 Yuen:1986PLA}.
In both scenarios, linearity of quantum mechanics forbids the existence of some 
 physical process, when the input set contains {\em all} states.
Even when the set of input states is restricted to two pure states, perfect 
 quantum cloning is impossible, as follows from unitarity.
It was furthermore shown that broadcasting (a natural generalization of quantum 
 cloning to mixed input states) is possible for a set of two mixed states, if 
 and only if the states commute \cite{Barnum:1996PRL}.
The same criterion does {\em not} apply for purification maps: a pair of 
 orthogonal or identical states can, of course, be purified perfectly -- but in 
 any other case of commuting states we will show that perfect purification is 
 impossible.
Yet for some non-commuting states, a perfect purification process exists.

\section{Two-state purifiers with pure output}\label{s4641}
In this section we will focus on the case of two input states and perfect 
 output purity, i.e. a deterministic process which takes any state from the set 
 $\M= \{\rho, \rho'\}$ to a pure state.
A characteristic quantity for purification will turn out to be the {\em 
 worst-case distinguishability} $\wcd\rho{\rho'}$, which denotes the trace 
 distance of the two closest states that may appear physically in the
 ensembles of $\rho$ and $\rho'$, i.e.
\begin{equation}\label{e22205}
 \wcd\rho{\rho'}= \min_{\ket\chi,\ket{\chi'}} \trnorm{\proj\chi- \proj{\chi'}},
\end{equation}
where $\ket\chi$ and $\ket{\chi'}$ are normalized vectors in the range of 
 $\rho$ and $\rho'$, respectively.
(We point out that this quantity can be calculated by taking the sine of the 
 smallest canonical angle \cite{Stewart:1990} between the range of $\rho$ and 
 the range of $\rho'$.)
The notion of distinguishability here refers to the success probability of a 
 minimum error discrimination, as explained above.

Although at first sight the worst-case distinguishability resembles a distance, 
 mathematically speaking it is none:
The triangular inequality does not hold, and $\wcd\rho{\rho'}=0$ is true for 
 some $\rho\ne \rho'$.
Note that any two states with overlapping ranges have, in fact, a vanishing 
 worst-case distinguishability.
On the other hand, $\wcd\rho{\rho'}=1$ is equivalent to $\rho$ and $\rho'$ 
 being orthogonal, i.e. $\trnorm{\rho- \rho'}=1$.
Thus commuting states are either orthogonal or have a vanishing worst-case 
 distinguishability.

\subsection{Characterization of two-state purifiers}\label{ss2943}
We are now in the position to study the general consequences of perfect output 
 purity.
Suppose that $\Lambda$ is a purifier of $\rho$ and $\rho'$ with perfect output 
 purity.
As a defining property of any normalized vector $\ket\chi$ in the range of 
 $\rho$ one can write $\rho=\alpha \proj \chi + \beta \tilde{\rho}$ with 
 positive numbers $\alpha$ and $\beta$, and positive semidefinite 
 $\tilde{\rho}$.
Using the same convexity argument as in the proof of Theorem 1 \textit{(i)}, it 
 follows that $\Lambda[\proj \chi]= \Lambda[\rho]$.
An analogous argument holds for all vectors $\ket{\chi'}$ in the range of 
 $\rho'$.
Thus we have $\trnorm{\Lambda[\rho]-\Lambda[\rho']} = 
 \trnorm{\Lambda[\proj\chi]- \Lambda[\proj{\chi'}]}\le \trnorm{\proj\chi- 
 \proj{\chi'}}$, where in the inequality we used that a deterministic physical 
 process $\Lambda$ cannot increase the trace distance between two states 
 \cite{Nielsen:2000}.
By choosing for $\proj\chi$ and $\proj{\chi'}$ the states with minimal distance 
 (cf. definition in Eq.~(\ref{e22205})), we have shown that for maps $\Lambda$ 
 where $\Lambda[\rho]$ as well as $\Lambda[\rho']$ are pure,
\begin{equation}\label{e20683}
 \wcd\rho{\rho'}\ge \trnorm{\Lambda[\rho]- \Lambda[\rho']}
\end{equation}
 must hold.

It is important that there always exists a map which reaches equality in 
 Eq.~(\ref{e20683}).
In order to see this, one constructs a canonical basis \cite{Stewart:1990} of 
 the ranges of both states, i.e. an orthonormal basis $\{\ket{\chi_i}\}$ of the 
 range of $\rho$ and $\{\ket{\chi'_i}\}$ of the range of $\rho'$, such that in 
 addition $\bracket{\chi_i}{\chi'_j}= 0$ holds for all $i\ne j$.
One can show that there always exists a map, which decreases the distance of 
 two pure states by an arbitrary value.
Such a map is now applied in each of the orthogonal subspaces spanned by 
 $\{\ket{\chi_i}, \ket{\chi'_i}\}$, such that the distance 
 $\trnorm{\proj{\chi_i}-\proj{\chi'_i}}$ decreases to be $\wcd\rho{\rho'}$.
The composed map has the property, that if applied to $\rho$ and $\rho'$, an 
 orthonormal eigenbasis for both output states exists, such that all 
 non-orthogonal eigenvectors (one of the output of $\rho$ and one of $\rho'$) 
 have a distance $\wcd\rho{\rho'}$.
Now a map can readily be found, which maps the output states to pure states 
 having a distance $\wcd\rho{\rho'}$.
The fact, that one can always reach the equality in Eq.~(\ref{e20683}), 
 completes the characterization of the output of a general process, which maps 
 two input states $\rho$ and $\rho'$ to two pure states.

\subsection{Bounds on two-state purifiers}
As an application of the result in Section~\ref{ss2943} we now estimate the 
 faithfulness of a purifier with perfect output in the case of two input 
 states.
For this purpose we assume that the state $\rho$ ($\rho'$) occurs with {\em a 
 priori} probability $\eta$ ($\eta'$), where $\eta'\geq \eta$ without loss of 
 generality.
We denote the deviation from perfect faithfulness by $\delta$, i.e.
\begin{equation}
 \delta= \eta\, \trnorm{\rho- \traux\Lambda[\rho]}+ \eta'\, \trnorm{\rho'-
          \traux\Lambda[\rho']}.
\end{equation}
Using the triangular inequality for the trace distance, $\trnorm{\rho- \rho'} 
 \le \trnorm{\rho- \traux\Lambda[\rho]}+ \trnorm{\traux\Lambda[\rho]- 
 \traux\Lambda[\rho']}+ \trnorm{\traux\Lambda[\rho']- \rho'}$ holds, and we 
 obtain due to Eq.~(\ref{e20683}) the lower bound
\begin{equation}\label{e25987}
 \delta\geq \eta\, [\trnorm{\rho- \rho'}- \wcd\rho{\rho'}].
\end{equation}

A straightforward upper bound on $\delta$ for the optimal process (i.e. minimal 
 $\delta$) can be obtained by considering a constant purifier that produces a 
 perfect purification of $\rho'$.
This leads to the first upper bound
\begin{equation}\label{e28966}
 \delta_\mathrm{opt}\leq \eta\, \trnorm{\rho- \rho'}.
\end{equation}
A more sophisticated upper bound on $\delta$ is given by using the map which 
 reaches the equality in Eq.~(\ref{e20683}).
One chooses the output of $\rho'$ to be a purification of $\rho'$ and the 
 output of $\rho$ to be a pure state, which is as close as possible -- 
 according to Eq.~(\ref{e20683}) -- to a purification of $\rho$.
Since the maximal overlap of all purifications for two states $\rho$ and 
 $\rho'$ is given by the Uhlmann fidelity $F(\rho,\rho')= \tr \sqrt{\sqrt\rho\, 
 \rho'\sqrt\rho}$ \cite{Uhlmann:1976RMP, Jozsa:1994JMO}, we find with 
 $\sin\alpha= \wcd\rho{\rho'}$ and $\cos\beta= F(\rho,\rho')$ the second upper 
 bound
\begin{equation}\label{e6944}
 \delta_\mathrm{opt}\leq \eta \sin(\beta-\alpha).
\end{equation}

\begin{figure}
 \resizebox{\columnwidth}{!}{\includegraphics{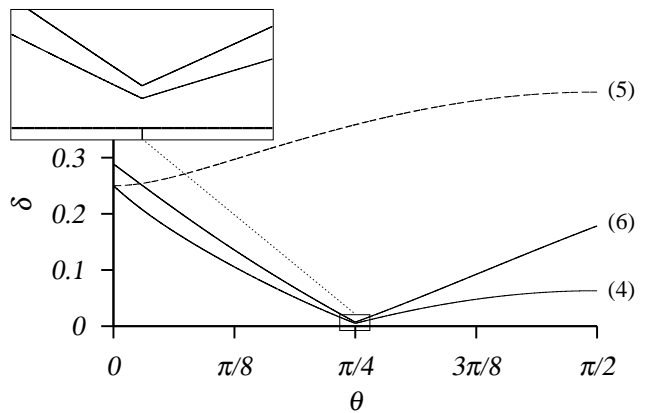}}
 \caption{
  Example for lower and upper bounds on the optimal deviation from perfect 
   faithfulness $\delta$ of a two state purifier with pure output.
  See main text for explanation.
 }
 \label{f2491}
\end{figure}

Let us give an explicit example for these bounds.
We consider the states $\rho= \frac 1 2(\proj 0+ \proj 1) \otimes \proj 0$ and 
 $\rho'= \frac 2 3 \proj 0 \otimes \proj{+}+ \frac 1 3 \proj 1 \otimes \proj 
 \theta$, which appear with equal {\em a priori} probability, where 
 $\ket\theta= \cos \theta \ket{0}+ \sin \theta \ket{1}$ and $\ket+= \frac 
 1{\sqrt2}(\ket 0+\ket 1)$.
In Fig.~\ref{f2491} the bounds for the optimal deviation from faithfulness 
 $\delta$ are shown: the lower bound as given in Eq.~(\ref{e25987}), the first 
 (dashed line) and second upper bound, cf. Eq.~(\ref{e28966}) and 
 (\ref{e6944}).
At $\theta= 0$ the ranges of both states share the vector $\ket 1\otimes \ket 
 0$ and thus the worst-case distinguishability vanishes and the optimal 
 faithfulness is given by the upper bound in Eq.~(\ref{e28966}).
The second upper bound and the lower bound almost coincide at $\theta= \pi/4$ 
 with $.0050<\delta<.0072$.
Note that the upper bounds cross each other, i.e. depending on the input state, 
 either the first or the second upper bound is tighter.

An interesting question in this context is the following: given two quantum 
 states, does a better distinguishability (in the sense of minimum error 
 discrimination) imply a better faithfulness?
The surprising answer is no: in the example given above, the trace distance of 
 the two states monotonically increases from $\theta= 0$ to $\theta= \pi/2$, 
 while the deviation from faithfulness has its minimum at $\theta= \pi/4$.
The examples illustrates, that the worst-case distinguishability is indeed an 
 important quantity for purifying processes.
This is remarkable, as the worst-case distinguishability is purely determined 
 by the geometric features of the states, whereas the statistical weights in 
 the ensembles do not play any role.
Note that a related, but not purely geometric quantity $F^+_1(\rho,\rho')$ was 
 introduced in \cite{Uhlmann:2000RMP}.

\section{Sets that can be purified perfectly}\label{s24871}
Finally, our focus turns to the general analysis of perfect purifiers.
The existence of a perfect purifier for a set $\M$ has far-reaching 
 implications, as it is possible to convert all states in $\M$ to pure states 
 in a reversible way.
An investigation of the property of reversibility indeed turns out to be the 
 key for understanding perfect purification:
Suppose that we have a purifier $\Lambda$ of a set $\M$ with perfect output 
 purity (but not necessarily perfect faithfulness), and some completely 
 positive and trace preserving map $\Lambda'$, such that for any $\rho_i \in 
 \M$ this map is the reverse map of $\Lambda$, i.e. 
 $\Lambda'[\Lambda[\rho_i]\,]= \rho_i$.
The action of any completely positive and trace preserving map can always be 
 formulated as appending a (pure) ancilla state, performing a unitary rotation 
 and finally tracing out an appropriate subsystem.
We write $\Lambda'$ in this manner and apply everything, apart from tracing 
 out, to the output of $\Lambda$.
For this composed map we write the shorthand notation $\tilde \Lambda$.
The output of $\tilde \Lambda$ is still pure for any state in $\M$ and the 
 remaining step of the map $\Lambda'$, namely the trace over the subsystem, 
 yields back the original state, thus $\tilde \Lambda$ is a {\em perfect} 
 purifier of $\M$.

In order to further approach the characterization of sets that can be purified 
 perfectly, we call a set of states {\em essentially pure}, if every state from 
 the set can be globally rotated into a tensor product of a pure state and a 
 common mixed contribution, or in more technical terms:
A set of states $\M$ is called essentially pure, if one can find states 
 $\omega_\mathrm{aux}$ and $\sigma_\mathrm{B}$, a unitary transformation $U$, 
 and a set of {\em pure} states $\mathcal P_\mathrm{A}$, such that for all 
 $\rho_i\in \M$ there is a corresponding pure state $\proj{\phi_i}\in \mathcal 
 P_\mathrm{A}$ with
\begin{equation}\label{e5983}
 \rho_i\otimes \omega_\mathrm{aux}= U(\proj{\phi_i}\otimes 
     \sigma_\mathrm{B})U^\dag.
\end{equation}
Note, that the tensor product symbol on the two sides of this equation in 
 general denotes {\em different} splits of the composite system: on the left 
 hand side one sees the composition of the original system and an auxiliary 
 system, while on the right hand side the composition refers to some system A 
 and some system B.
Essentially pure sets can be purified perfectly:
A process which appends $\omega_\mathrm{aux}$ to $\rho_i$, performs $U^\dag$ 
 and traces out system B produces a pure state for any state in $\M$.
On the other hand a process, which appends $\sigma_\mathrm{B}$ to 
 $\proj{\phi_i}$, performs $U$ and traces out the auxiliary system, undoes the 
 action of the purifying map.
Thus, a perfect purifier of $\M$ exists.
Of course a union of essentially pure sets, where any two states taken from 
 different sets are orthogonal, can also be purified perfectly.
We call such a union an {\em orthogonal union} of essentially pure sets.

\begin{theorem}\label{t21}
 For a set of states $\M$, the following statements are equivalent:
 (i)~A perfect purifier of $\M$ exists.
 (ii)~There exists a completely positive and trace preserving map, which maps 
  any state in $\M$ to a pure state and does not change the trace distance of 
  any two states in $\M$.
 (iii)~$\M$ is an orthogonal union of essentially pure sets.
\end{theorem}
\begin{proof}
 Our motivation for the definition of orthogonal unions of essentially pure 
  sets was indeed, that this property implies the existence of a perfect 
  purifier.
 Thus, we have already shown that \textit{(iii)} implies \textit{(i)}.
 Furthermore, from the fact that no process can increase the trace distance, 
  together with the existence of a reversible map, \textit{(ii)} is a direct 
  consequence of \textit{(i)}.
 Thus it only remains to show that \textit{(ii)} implies \textit{(iii)}:
 If \textit{(ii)} holds for an  $\M$ that is a union of mutually orthogonal
  subsets, there exist maps that satisfy \textit{(ii)} for each subset.
 Therefore, we can assume without loss of generality that one cannot  split the 
  set $\M$ into orthogonal parts.
 With $\proj a$ being a pure auxiliary state and $U^\dag$ a unitary 
  transformation, we can write the action of $\Lambda$ as $\rho\mapsto 
  \tr_\mathrm{B} U^\dag(\rho\otimes \proj a)U$, where B denotes an appropriate 
  subsystem.
 Since the output of $\Lambda$ for a state $\rho_i \in \M$ is a pure state 
  (represented by a projector $\Phi_i$), we have $U^\dag(\rho_i\otimes \proj 
  a)U= \Phi_i\otimes \sigma_i$, with $\sigma_i$ a state in subsystem $\mathrm 
  B$.
 The final step is now to show that $\sigma_i= \sigma_j$ holds.
 For any two states $\rho_i, \rho_j\in \M$, due to the assumption 
  \textit{(ii)},
\begin{equation}\label{e11294}
 \trnorm{\Phi_i-\Phi_j}= \trnorm{\rho_i- \rho_j}=
  \trnorm{\Phi_i\otimes \sigma_i- \Phi_j\otimes \sigma_j}
\end{equation}
  holds.
 A minimum error discrimination \cite{Helstrom:1976, Herzog:2004PRA} in 
  subsystem B at the right hand side can be written as $\sigma_i\rightarrow q_i 
  \proj 0+ (1-q_i) \proj 1$ and $\sigma_j \rightarrow (1- q_j) \proj 0+ q_j 
  \proj 1$, where $(q_i+ q_j)/2=(1+ \trnorm{\sigma_i- \sigma_j})/2$ is the 
  success probability for the optimal discrimination measurement.
 We find
\begin{equation}\begin{split}
  \|&\Phi_i\otimes \sigma_i -\Phi_j\otimes \sigma_j\|^2
 \\& \ge (
   \trnorm{q_i\Phi_i- (1- q_j)\Phi_j}+ \trnorm{(1- q_i)\Phi_i- q_j\Phi_j})^2
 \\& \ge
  \trnorm{\Phi_i- \Phi_j}^2+
  \trnorm{\sigma_i- \sigma_j}^2 \,\tr(\Phi_i \Phi_j),
\end{split}\end{equation}
  where in the first step we used, that the discrimination procedure cannot 
  increase the trace distance.
 The second inequality follows from a lengthy but straightforward calculation.
 From comparison with Eq.~(\ref{e11294}) either $\sigma_i= \sigma_j$ or 
  $\tr(\Phi_i\Phi_j)= 0$ (or both) must hold.
 The latter case implies $\rho_i$ to be orthogonal to $\rho_j$, i.e. if 
  $\sigma_i\ne \sigma_j$ for two states, then one can split $\M$ into two 
  orthogonal sets, in contrast to our assumption.
\end{proof}

This Theorem completely characterizes all sets of states that can be purified 
 perfectly, cf. also Eq.~(\ref{e5983}).
It is surprising that one can even purify a set of {\em continuous} states,
 meaning that the set may contain infinitesimally close neighbors.
It is also worth mentioning that all states in an essentially pure set share 
 the same spectrum and pairwise have a completely degenerate set of canonical 
 angles \cite{Stewart:1990}.
What is the lowest dimension, in which perfect purification is possible for 
 nonorthogonal mixed states?
This cannot happen unless the dimension of the Hilbert space is at least four:
In two and three dimensions, only pure states can have identical spectra 
 without having an overlapping range.

Although essentially pure sets can be characterized in a explicit manner and 
 have a lot of straightforward features, there is no obvious method to verify 
 whether a given set is of the structure as specified in Eq.~(\ref{e5983}).
However, for the case, where $\M$ consists of only two states, there exists a 
 computable test:
From the lower bound on $\delta$ derived in equation (\ref{e25987}) it follows 
 that $\trnorm{\rho- \rho'}= \wcd\rho{\rho'}$ is a necessary condition for the 
 existence of a perfect two-state purifier.
It is also a sufficient condition:
For any two states $\rho$ and $\rho'$ there is a map $\Lambda$ such that 
 $\trnorm{\Lambda[\rho]- \Lambda[\rho']}= \wcd\rho{\rho'}$, thus if 
 $\trnorm{\rho- \rho'}= \wcd\rho{\rho'}$, this map satisfies part \textit{(ii)} 
 of Theorem~\ref{t21}, i.e. $\rho$ and $\rho'$ can be purified perfectly.
Note, that it is also straightforward to prove that the upper bound on 
 $\delta_\mathrm{opt}$ in Eq.~(\ref{e6944}) vanishes if and only if there is a 
 perfect purifier of $\rho$ and $\rho'$.

\section{Conclusions}
In summary, we have introduced the concept of purification as a physical map, 
 and studied its properties: without any prior knowledge of the input state a 
 perfect purifier cannot exist.
Relaxing one of the two characteristic properties of a purifier, purity and 
 faithfulness, does not lead to a non-trivial universal process either.
We have investigated the case when the input set contains only two states and 
 found a characterization of the output of any map, which takes both states to 
 a pure state.
Using this tool, we derived bounds on the deviation from perfect faithfulness 
 (i.e. the distance of the partial trace of the output state and the original 
 state).
We also completely characterized all sets of states, that can be purified 
 perfectly.
Roughly speaking, any such set can be globally rotated into a set of pure 
 states, tensored with a common mixed contribution.
Surprisingly, we found that some sets of non-commuting states can be purified, 
 in contrast to the situation of broadcasting.
For the case of sets with only two states, we provided an operational test to 
 check whether perfect purification is possible.

In this paper we have presented some of the basic properties of purifying 
 completely positive maps.
Several questions remain open.
One direction of future work is to consider the maximal possible purity of a 
 purifier in the case of perfect faithfulness.
Furthermore, the analysis of purifiers for sets with more than two states will 
 be subject of further research.

\acknowledgements
We acknowledge discussions with Norbert L\"utkenhaus, Armin Uhlmann and 
 Reinhard Werner.
This work was partially supported by the EC programmes SECOQC and SCALA.

\bibliography{the}
\end{document}